\documentclass[aps,prb,showpacs]{revtex4}
\usepackage{epsfig,amsfonts,amssymb}
\begin{document}
\title{Geometric phases and quantum entanglement as building blocks
\\for nonabelian quasiparticle statistics}
\author{Ady Stern}
\affiliation{Department of Condensed Matter Physics,
    Weizmann Institute of Science, Rehovot 76100, Israel}
\author{Felix von Oppen}
\affiliation{Department of Condensed Matter Physics,
    Weizmann Institute of Science, Rehovot 76100, Israel\\
    Institut f\"ur Theoretische Physik, Freie Universit\"at Berlin,
Arnimallee
14, 14195 Berlin, Germany\footnote{Permanent address}}
\author{Eros Mariani}
\affiliation{I.\ Institut f\"ur Theoretische Physik, Universit\"at
Hamburg, Jungiusstr.\
9, 20355 Hamburg, Germany\\
Department of Condensed Matter Physics,
    Weizmann Institute of Science, Rehovot 76100, Israel$^*$}
\date{\today}
\begin{abstract}
Some models describing unconventional fractional quantum Hall
states predict quasiparticles that obey nonabelian quantum
statistics. The most prominent example is the Moore-Read model for
the $\nu=5/2$ state, in which the ground state is a superconductor
of composite fermions, the charged excitations are vortices in
that superconductor, and the nonabelian statistics is closely
linked to the degeneracy of the ground state in the presence of
vortices. In this paper we develop a physical picture of the
nonabelian statistics of these vortices. Considering first the
positions of the vortices as fixed, we define a set of
single-particle states at and near the core of each vortex, and
employ general properties of the corresponding Bogolubov-deGennes
equations to write the ground states in the Fock space defined by
these single-particle states. We find all ground states to be {\it
entangled superpositions} of all possible occupations of the
single-particle states near the vortex cores, in which the
probability for all occupations is equal, and the relative phases
vary from one ground state to another. Then, we examine the
evolution of the ground states as the positions of the vortices
are braided. We find that as vortices move, they accumulate a {\it
geometric phase} that depends on the occupations of the
single-particle states near the cores of other vortices. Thus,
braiding of vortices changes the relative phase between different
components of a superposition, in which the occupations of these
states differ, and hence transform the system from one ground
state to another.  These transformations, that emanate from the
quantum entanglement of the occupations of single-particle states
and from the dependence of the geometric phase on these
occupations, are the source of the nonabelian statistics. Finally,
by exploring a ``self-similar" form of the many-body wave
functions of the various ground states, we show the equivalence of
our picture, in which vortex braiding leads to a change in the
relative phase of components in a superposition, and pictures
derived previously, in which vortex braiding seemingly affects the
occupations of states in the cores of the vortices.

\end{abstract}

\maketitle

\section{Introduction}

The experimental discovery \cite{stormer-fqhe} of the fractional
quantum Hall effect (FQHE) led to intriguing theoretical
observations regarding the elementary excitations (quasiparticles)
of a two-dimensional electron system at a fractional Landau-level
filling factor $\nu $. Very soon after the experimental discovery,
Laughlin \cite{Laughlin} realized that the quasiparticles at
filling factors $\nu =1/(2p+1)$ (with $p$ an integer) carry a
fractional charge $e^{\ast }=\pm e/(2p+1)$ (for brevity, we use
the term quasiparticles to refer also to quasiholes). Following
that observation, Halperin \cite{Halperin} showed that the
hierarchy of observed FQHE states, at $\nu =p/q$ (with $q$ an odd
integer), points to the fractional statistics of the
quasiparticles and quasiholes, of the type that was previously
studied by Wilczek. \cite{Wilczek} This observation was further 
clarified by Arovas, Schrieffer, and Wilczek. \cite{Arovasetal}
When a system contains two quasiparticles, and the positions of
these quasiparticles are adiabatically interchanged, the state of
the system acquires a geometric Berry phase. This phase, which is
$\pi $ for fermions and $2\pi $ for bosons, becomes a fraction of
$\pi $ for FQHE quasiparticles.

The experimental discovery \cite{Willett} of the even-denominator
FQHE state $\nu =5/2$ triggered the introduction of yet another
novel concept with regard to the statistics of the elementary
excitations. Employing conformal field theory to study the $\nu
=5/2$ FQHE, Moore and Read \cite{MooreRead} discovered that if
this state is well described by the Pfaffian wave function, as numerical investigations seem to confirm, \cite{Morf} the
elementary excitations obey nonabelian statistics. The state of
the system after a series of quasiparticle interchanges then
depends on the order in which these interchanges are carried out.
By using exact eigenstates of a model Hamiltonian, \cite{Greiter}
Nayak and Wilczek \cite{NayakWilczek} subsequently showed that the
ground state of the configuration in which $2N$ quasiholes are
inserted at fixed positions is $2^N$-fold degenerate, and that the
quasiparticles realize a $2^{N-1}$-dimensional spinor braiding
statistics. Following earlier observations that related the
Pfaffian state to $p-$wave Cooper-pairing, Read and Green
\cite{ReadGreen} described this state as a $p$-wave BCS
superconductor of composite fermions, and conjectured that the
nonabelian statistics of its quasiparticles results from the
zero-energy modes associated with vortices in this superconductor.

This conjecture was further studied by Ivanov, \cite{Ivanov} who
mapped out the relation between the exchange of quasiparticles and
the unitary transformation carried out on the Hilbert space of the
ground states. Ivanov explicitly derived these unitary
transformations, showed that they are indeed nonabelian, and
confirmed that they are identical to the transformations derived
earlier by Nayak and Wilczek \cite{NayakWilczek} using conformal
field theory.

While these two derivations of the unitary transformations
associated with vortex interchange may be easily generalized to
calculate the transformations associated with other braidings of
vortices, they do not provide a clear physical picture of the
nonabelian statistics. This is exemplified in the following
observation: using the method of Ref.\ \onlinecite{Ivanov}, it is
easy to show that when the system is initially in a ground state
$\left| {\rm gs}_{\alpha }\right\rangle $, and vortex $j$
encircles vortex $j+1,$ then, {\it under the assumption that no
tunneling takes place between the vortex cores}, the final state
of the system is again a ground state, given by
\begin{equation}
\left( c_{j}e^{\frac{i}{2}\Omega
_{j}}+c_{j}^{\dagger}e^{-\frac{i}{2} \Omega _{j}}\right) \left(
c_{j+1}e^{\frac{i}{2}\Omega _{j+1}}+c_{j+1}^{\dagger}e^{-\frac{
i}{2}\Omega _{j+1}}\right) \left| {\rm gs}_{\alpha }\right\rangle,
\label{unit-trans}
\end{equation}
where the operators $c_{j}^{(\dagger)},c_{j+1}^{(\dagger)}$
annihilate (create) a particle localized very close to the cores
of the $j$th and $(j+1)$th vortex, respectively, and $\Omega_j$ is
a  phase defined in the next section. Eq.\
(\ref{unit-trans}) seemingly implies that the motion of the $j$th
vortex around the $(j+1)$th vortex affects the occupations of
states very close to the cores of the two vortices. This is
in contrast, however, to the derivation leading to Eq.\ (\ref
{unit-trans}), which explicitly assumes that vortices are kept far
enough from one another so that tunneling between vortex cores may
be disregarded.

In this work we study the ground states $\left|{\rm gs}_{\alpha }
\right\rangle $ in an attempt to give a physical picture of the effect
of braidings in the positions of vortices. We show that two
ingredients are essential for the nonabelian statistics of the
vortices. The first is the {\it quantum entanglement} of the
occupation of states near the cores of distant vortices. The second
ingredient is familiar from (abelian) fractional statistics: the {\it
  geometric phase} accumulated by a vortex traversing a closed loop.

Within the Chern-Simons composite-boson theory, \cite{SCZhang} the abelian
fractional statistics of the $\nu=1/m$ states is explained by mapping
the ground state of the electronic system to a superfluid of composite
bosons, and the quasiparticle excitations to vortices in that
superfluid. Due to the coupling of the vortex to a Chern-Simons gauge
field, the depletion of bosons at the vortex core is quantized to a
fraction $1/m$ of a fluid particle. Thus the charge carried by the
vortex is also fractional. The quantum statistics is related to the
geometric phase accumulated by a vortex traversing a close trajectory.
Roughly speaking, the vortex accumulates a phase of $2\pi$ per fluid
particle which it encircles.  When another vortex with its fractional
charge is introduced to the encircled area, this phase changes by a
fraction of $2\pi$.  Upon adapting the argument to interchanging of
vortices, one finds that this fraction of $2\pi$ translates into
fractional statistics.

Similarly, the Moore-Read theory of the $\nu=5/2$ state describes it also as a
superfluid, with the quasiparticles being vortices in that superfluid.
However, the ``effective bosons'' forming the superfluid are Cooper
pairs of composite fermions. Consequently, the superfluid has
excitation modes associated with the breaking of Cooper-pairs.  In the
presence of vortices, a Cooper-pair may be broken such that one or two
of its constituents are localized in the cores of vortices. For
$p-$wave superconductors, the existence of zero-energy intra-vortex
modes leads, first, to a multitude of ground states, and, second, to a
particle-hole symmetric occupation of the vortex cores in all ground
states. When represented in occupation-number basis, a ground state is
a superposition which has equal probability for the vortex core being
empty or occupied by one fermion.

When a vortex traverses a trajectory that encircles another vortex,
the phase it accumulates depends again on the number of fluid
particles which it encircles. Since a fluid particle is, in this case, a
Cooper pair, the occupation of a vortex core by a fermion, half a
pair, leads to an accumulation of a phase of $\pi$ relative to the
case when the core is empty. And since the ground state is a
superposition with equal weights for the two possibilities, the
relative phase of $\pi$ introduced by the encircling might in this
case transform the system from one ground state to another.

This qualitative picture is made more precise in this paper. Our
analysis revolves around the definition of a set of single-particle
states localized at or near the $2N$ vortex cores. We start by
defining the ``core states," a set of $2N$ states each of which is
localized at a specific vortex core. We find that in all possible
ground states, the occupation of these single-particle states near one
vortex is entangled with the occupation of single-particle states near
all other vortices. We prove that any many-body state in which these
occupations are disentangled is necessarily an excited state. We show
that the evolution of the ground state as positions of vortices are
braided indeed follows the picture outlined above, and discuss both the
case where vortices encircle one another and the case where they
interchange positions. In making this picture of nonabelian
statistics more precise, we define, starting from each core state,
further orthogonal single-particle states (``near-core states") which
are localized near a vortex core. The occupations of these additional
single-particle states in the many-body ground states are also
particle-hole symmetric and entangled between different vortices. In
fact, we reveal a ``self-similar" structure of the many-body
wave function with respect to the occupation of these single--particle
states which leads us to express the relation between our picture of
nonabelian statistics and the known representations of vortex braiding
in the space of ground states \cite{NayakWilczek,Ivanov} in terms of
compact operator identities.

The structure of this paper is as follows: We begin in Sec.\
\ref{bdg-section} with a review of the description of the Pfaffian
state as a $p-$wave superconductor, with vortices as quasiparticles.
In Sec.\ \ref{core-section}, we start with the definition of the
single-particle states by introducing the ``core states." In Sec.\
\ref{entanglement-section}, we explore the roles of quantum
entanglement and geometric phases in the evolution of these
superpositions when vortex positions are braided. The ``near-core states"
are introduced in Sec.\ \ref{nearcore-section}. The occupations of
these ``near-core states" in the many-body ground states is worked out
in Sec.\ \ref{near-core-wf-section}, revealing the ``self-similar"
structure of the wave functions. We conclude in Sec.\
\ref{conclusions}. Some details are relegated to appendices.

\section{Solutions of Bogolubov-deGennes equations - review}
\label{bdg-section}

The Pfaffian trial wave function for the quantized Hall state at
Landau level filling factor $\nu =5/2$ was first introduced by Moore
and Read\cite{MooreRead} as the first-quantized wave function
\begin{equation}\label{Pfaffian}
\Psi_{\mathrm{MR}}^{}(z_{1},z_2,...) =  \textrm{Pf}\left
(\frac{1}{z_{i}^{}-z_{j}^{}}\right)\prod_{i<j}\left(z_{i}-z_{j}
\right)^{2} \prod_{j}e^{-\frac{1}{4 \ell^{2}_{}}|z_{j}|^{2}},
\end{equation}
where $\ell$ is the magnetic
length and $z_{i}=x_i+iy_i$ is the complex coordinate of the $i$th
particle. For $p$ particles, the Pfaffian in Eq.\ (\ref{Pfaffian})
takes the explicit form
\begin{equation}\label{Pf}
\textrm{Pf} \left({1\over z_i-z_j}\right) = {1\over 2^{p/2}(p/2)!}\,{\cal A}  \left\{
{1\over z_1-z_2}{1\over z_3-z_4}\cdots {1\over z_{p-1}-z_p}\right\},
\end{equation}
where ${\cal A}$ is the antisymmetrization operator.
It is instructive to view the
Pfaffian appearing in the wave function in Eq.\ (\ref{Pfaffian}) as
the real-space BCS wave function of composite
fermions for a fixed number of particles. \cite{ReadGreen,Schrieffer}
According to the associated pair wave function $g(z) = 1 / z $, the
pairing is of spinless (or spin-polarized) composite fermions in the
$l=-1$ angular-momentum ($p$-wave) channel. The Pfaffian corresponds
to a weakly-paired superconductor, which for a
two-dimensional $p$-wave superconductor is topologically distinct from the
strongly-paired phase. \cite{ReadGreen} The charged excitations of the
quantum-Hall system are, in this description, the half-flux-quantum
($h/2e$) vortices of the superconductor.

As is often the case, the use of a first-quantization formulation is
mathematically involved, and makes the physical picture difficult to
read. The identification of the Pfaffian as a complex $p$-wave BCS
state of composite fermions subsequently led Read and Green
\cite{ReadGreen} to introduce a second-quantization formulation which
paved the way for a clearer physical picture. Their starting point is
the BCS mean field Hamiltonian
\begin{equation}
H=\int \mathrm{d}{\bf r}\,\psi ^{\dagger }({\bf r})h_{0}\psi ({\bf
r})+ {\frac{1}{2}} \int \mathrm{d}{\bf r}\,\mathrm{d}{\bf
r^{\prime }}\left\{ D^{\ast }({\bf r},{\bf r^ {\prime
}})\psi ({\bf r^{\prime }})\psi ({\bf r})+D({\bf r},{\bf
r^{\prime }})\psi ^ {\dagger }({\bf r})\psi ^{\dagger }({\bf
r^{\prime }} )\right\}  \label{bcs-hamiltonian}
\end{equation}
with the single-particle term $h_0$ and the complex $p$-wave pairing
function
\begin{equation}
D({\bf r},{\bf r}^{\prime })=\Delta \left( {\frac{{\bf r}+{\bf r^
{\prime
}}}{2}}\right) (i\partial _{x^{\prime }}-\partial _{y^{\prime }})\delta
({\bf r}-{\bf
r^{\prime }}).
\end{equation}
Read and Green \cite{ReadGreen} retain only the potential part of $h_{0}
$ by setting
$h_{0}=-\mu ({\bf r})$ and argue that this is sufficient for studying
the topological
properties of the Pfaffian state, such as the statistics of its
quasiparticles. In the
presence of $2N$ vortices pinned at positions ${\bf R} _i$, the gap
function takes the
form $\Delta({\bf r}) = |\Delta({\bf r})| \exp[i\chi({\bf r})]$ with
$\chi({\bf r}) =
\sum_{i=1}^{2N}\arg({\bf r}-{\bf R}_i)$. In the vicinity of vortex $k$,
the phase
$\chi({\bf r})$ can be approximated by $\chi({\bf r}) = \arg({\bf r}-
{\bf R}_k) +
\Omega_k$ with $ \Omega_k = \sum_{i\neq k}^{2N}\arg({\bf R}_k-{\bf R}_i)
$.

The fermionic excitations of superconductors are described by the
Bogolubov-deGennes (BdG) equations
\begin{equation}
E\left(
\begin{array}{c}
u({\bf r}) \\
v({\bf r})
\end{array}
\right) =\left(
\begin{array}{cc}
-\mu ({\bf r}) & \frac{i}{2}\left\{ \Delta ({\bf r}),\partial _{x}
+i\partial
_{y}\right\} \\
\frac{i}{2}\left\{ \Delta ^{\ast }({\bf r}),\partial _{x}-i\partial
_{y}\right\} & \mu ({\bf r})
\end{array}
\right) \left(
\begin{array}{c}
u({\bf r}) \\
v({\bf r})
\end{array}
\right).
\end{equation}
For two-dimensional complex $p$-wave superconductors, solutions of non-zero
energy should be distinguished from those of zero energy. We
denote the non-zero energy solutions by $(u_{E}({\bf r}),v_{E}({\bf
  r}))$, with $u_{E}^{}({\bf r})=v_{-E}^{*} ({\bf r})$. In second
quantization, positive-energy solutions are associated with
annihilation operators of BCS quasiparticles $\Gamma _ {E}=\int
\mathrm{d}{\bf r}\left[ u_{E}({\bf r})\psi ({\bf r})+v_{E}({\bf
    r})\psi ^{\dagger}({\bf r}) \right] $, while negative-energy
solutions are associated with creation operators of the same
quasiparticles, $\Gamma _{-E}=\Gamma _{E}^{\dagger}$. The zero-energy
solutions $(u_{i}({\bf r}),v_{i}({\bf r}))$ are localized at the
vortex cores.  For well-separated vortices, there is one such solution
per vortex. With a choice of phase the advantages of which become clear
below, these zero-energy solutions take the form
\begin{equation}
u_{i}({\bf r})=v_{i}^{\ast }({\bf r} )= \frac{1}{\sqrt{2}}\, w_{i}^{(0)}
({\bf
r})\,e^{\frac{i}{2}\Omega _{i}}. \label{zeromodespinor}
\end{equation}
Here, the index $i=1,\ldots, 2N$ numbers the vortices and the functions
$w_{i}^{(0)}({\bf
r})$ are normalized wave functions localized near the core of the $i$th
vortex. When the
vortices are well separated, the functions $w_{i}^{(0)}({\bf r})$ are
mutually
orthogonal. In Sec.\ \ref{nearcore-section} below, we iteratively define
additional single-particle states
localized in the vicinity of the vortex cores. These
states will be
denoted by $w_{j}^{(k)}({\bf r})$, with the
subscript
labelling the vortex and the superscript enumerating the states near
each vortex.

The zero-energy eigenstates of the BdG equations correspond
to the Bogolubov
operators
\begin{equation}
\gamma _{j}={\frac{1}{\sqrt{2}}}\left[c_{j}e^{\frac{i}{2}\Omega
_{j}}+c_{j}^{\dagger}e^{-\frac{i}{2}\Omega _{j}}\right],  \label{gamma}
\end{equation}
where we introduced the operators $c_{j}=\int \mathrm{d}{\bf r}\,w_{j}^
{(0)}( {\bf r})\psi ({\bf
r})$ which create particles in the vortex-core states $ w_{j}^{(0)}
({\bf r})$.
Evidently,
$\gamma _{j}^{\dagger}=\gamma _{j}$ so that the operators associated
with the zero-energy
solutions are Majorana fermions.

The existence of the zero-energy solutions leads to a degeneracy of the
ground state. Enumeration of the ground states is
customarily done by combining the Majorana
operators $\gamma_{i}$ in pairs and defining (ordinary) fermionic creation and
annihilation operators
\begin{equation}
\alpha_{2j}^\dagger=\frac{1}{\sqrt{2}} \left( \gamma _{2j-1}+ i\gamma
_{2j}\right)\label{alpha}
\end{equation}
with $j=1,\ldots, N$.
The ground states can now be written in the occupation-number basis
corresponding to these
fermionic operators. A ground state $|{\bf m}\rangle$ is then labelled
by the occupation numbers ${\bf
m}=\left( m_{2,}m_{4},\ldots, m_{2N}\right)$ with
\begin{eqnarray}
    \alpha_{2j}^\dagger\alpha_{2j} |{\bf m}\rangle = m_{2j}|{\bf m}
\rangle
    \label{occupationofalpha}
\end{eqnarray}
and
\begin{eqnarray}
   |{\bf m}\rangle = (\alpha_{2N}^\dagger)^{m_{2N}}\ldots(\alpha_{4}
^\dagger)^{m_4}
   (\alpha_{2}^\dagger)^{m_2} |{\bf m}=0\rangle
\end{eqnarray}
leading to a $2^N$-fold degeneracy of the ground state.

The BCS Hamiltonian Eq.\ (\ref{bcs-hamiltonian}) is diagonal when
written in terms of the quasiparticle operators $\Gamma _{E}$ and
$\gamma _{i}$ (Bogolubov transformation). The ground state is
determined by the conditions $\Gamma _{E}\left| {\rm gs} \right\rangle
=0$ for all $E>0$. For a uniform superconductor in the absence of
vortices (i.e., for space-independent $\Delta $ and $ \mu $), these
equations lead to the celebrated BCS wave function $\left| {\rm
    BCS}\right\rangle =\prod_{\mathbf{k}}^{\prime}\left (
  u_{\mathbf{k}}+v_{\mathbf{k}}c_{\mathbf{k}}^{\dagger}c_{-\mathbf{k}}^
  {\dagger}\right) \left| {\rm vac} \right\rangle $, where the prime
indicates that pairs of momenta $(\mathbf{k},-\mathbf{k})$ should be
counted only once, and $|{\rm vac}\rangle$ is the state with no particles. In this case, the choice of a plane-wave basis for
the single-particle states is natural.

In the presence of vortices, $\Delta $ and $\mu $ are space dependent,
and there is no obvious single-particle basis for the description of
the ground states.  A proper choice of such a
basis turns out to be helpful in our discussion of nonabelian
statistics.

\section{Core states\label{core-section}}

A natural starting point for a single-particle basis are the $2N$
states $ w_{i}^{(0)}({\bf r}).$ These single-particle states, which
we approximate to
be orthogonal due to the large distance between the vortices, are
associated with a $2^{2N}$-dimensional subspace of the many-particle
states. The remaining (infinitely-many) single-particle basis states
remain unspecified throughout this section and are partially defined
in Sec.\ \ref{nearcore-section}. We will refer to the first $2N$
single-particle states as the ``vortex-core states,'' and to the remaining
single-particle states as
the ``other states.''

Many-particle states in Fock space can now be expanded in terms of
occupation numbers of these single-particle states. A corresponding
basis state of the Fock space is then written as
\begin{eqnarray}
&&\underbrace{\left| 1\ldots \ldots \ldots 0\right\rangle }\underbrace{
\left| 0\ldots 1\right\rangle } ,  \nonumber \\
&&{\rm vortex\,\,\,core}\,\,\,\,\,\,\,\,\,{\rm other}
\label{basis-states}
\end{eqnarray}
where $0$ ($1$) denotes an empty (occupied) state.
The first factor has $2N$ digits and designates the occupations of the
vortex-core
states. We enumerate these states by $\left| \tau \right\rangle $ with
$\tau =1,\ldots,
2^{2N}$. The second factor in Eq.\ (\ref {basis-states}) designates the
occupations of
the ``other states''.
Below we find that for every possible
ground state $\left| {\rm {gs}_{\alpha }}\right\rangle $ the
probability to find the core states in an occupation $\left| \tau
\right\rangle$ is equal to $1/2^{2N}$, and is independent of $\tau
$. A state in which this occupation depends on $\tau$ is
necessarily an excited state.

Although the operators $\alpha_{2j}$ and $\alpha_{2j}^\dagger$ act on the
occupations of the core states only, the quantum numbers ${\bf m}$ do not fully label the $2^{2N}-$dimensional space of states $\left |\tau\right\rangle$. Specifically, there is no direct relation between the occupation numbers ${\bf m}$ and the occupation of the single-particle states $w_j^{(0)}({\bf r})$. In order to explore the structure of the ground states in terms of the states $\left|\tau\right\rangle$, we now introduce
another set of quantum
numbers associated with a second set of $N$ Majorana operators,
\begin{equation}
X_{j}= {\frac{i}{\sqrt{2}}}\left( c_{j}e^{\frac{i}{2}\Omega
_{j}}-c_{j}^{\dagger}e^{-\frac{i}{2}\Omega _{j}}\right) \label
{xoperators}
\end{equation}
whose associated BdG spinors $(iw_{j}^{(0)}({\bf r})e^{{i\over
2}\Omega_j}\,,\,-i[w_{j}^{(0)}({\bf r})]^*e^{-{i\over
2}\Omega_j})$ are orthogonal by construction to the zero-energy
solutions of the BdG equations. These operators obviously also act
only on the occupations of the vortex-core states. However, when
expanded over the complete set of BdG quasiparticle operators,
they involve only non-zero energy quasiparticles so that
\begin{equation}
X_{j}=\sum_{E>0}\left[ C_{E}^{j}\Gamma _{E}+C_{E}^{j\ast }\Gamma _{E}^
{\dagger}\right]
\label{xdecomposed}
\end{equation}
with coefficients $C_{E}^{j}= i\sqrt{2}\int \mathrm{d}{\bf r}\, u^*_{E}
({\bf r} )w_{j}^{(0)}({\bf
r})e^{i\Omega _{j}/2}$. Upon pairing the vortices, these
Majorana operators can
again be combined to obtain (ordinary) fermionic operators\cite{foot1}
\begin{equation}
    \beta^\dagger_{2j} = {1\over \sqrt{2}} (iX_{2j-1} + X_{2j}).
\end{equation}
We label the occupation numbers of these fermions by  $ x_{2},x_
{4},\ldots,x_{2N},$ and
introduce ${\bf x}=\left(x_{2},x_{4}, \ldots,x_{2N}\right)$.

We now form a basis of the $2^{2N}$-dimensional Fock subspace of
the vortex-core states by defining
\begin{equation}
   |{\bf m},{\bf x}\rangle = (\alpha_{2N}^\dagger)^{m_{2N}}\ldots
(\alpha_{4}^\dagger)^{m_4}
   (\alpha_{2}^\dagger)^{m_2} (\beta_{2}^\dagger)^{x_{2}}(\beta_{4}
^\dagger)^{x_4}\cdots
   (\beta_{2N}^\dagger)^{x_{2N}} |{\bf m}=0,{\bf x}=0\rangle\label
{labelmx}
\end{equation}
where $|{\bf m}=0,{\bf x}=0\rangle$ is the state annihilated by
all the $\alpha_ {2j}$'s and $\beta_{2j}$'s. We obviously have
$\left\langle {\bf m},{\bf x}|{\bf m'},{\bf
x'}\right\rangle=\delta_{{\bf m}{\bf m'}}\delta_{{\bf x}{\bf
x'}}$. In terms of the occupations of the vortex-core states, the
states (\ref{labelmx}) take the explicit form \cite{foot2}
\begin{eqnarray}
\left| {\bf m},{\bf x}\right\rangle &=& s_{{\bf m},{\bf x}}
\prod_{j=1}^{N} \left\{ \left[ 1+i\,(-1)^{x_{2j}}
e^{-\frac{i}{2}\left( \Omega _{2j-1}+\Omega _{2j}\right)
}c_{2j-1}^{\dagger}c_{2j}^{\dagger}\right] \delta
_{m_{2j},x_{2j}}\right.  \nonumber \\
&&\left. +\left[ e^{-\frac{i}{2}\Omega _{2j-1}}c_{2j-1}^{\dagger}+i\,
(-1)^{x_{2j}}e^{-\frac{i}{2}\Omega _{2j}}c_{2j}^{\dagger}\right] \delta
_{m_{2j}+x_{2j},1}\right\} \left| {\rm vac}\right\rangle \; . \label
{cxstate}
\end{eqnarray}
The sign factor $s_{{\bf m},{\bf x}} =  \prod_{l=2}^N
\prod_{r=1}^{l-1}(-1)^{x_{2l}(m_{2r}+x_{2r})}$ arises due to the
different operator
orderings in Eq.\ (\ref{labelmx}) and in the product in Eq.\ (\ref
{cxstate}).

A ground state labelled by ${\bf m}$ is then a superposition
of the form
\begin{equation}
\left| {\bf m}\right\rangle =\sum_{{\bf x}} \left| {\bf m},{\bf x}
\right\rangle \left|
A_{{\bf x} }\right\rangle .  \label{cis}
\end{equation}
It is important to note that the states $|A_{\bf x}\rangle$ are {\it
independent} of the
particular ground state $|{\bf m}\rangle$.  Arbitrary ground states
$\left| {\rm
{gs}_{\alpha }}\right\rangle $ can be written as linear superpositions
of the states
$\left| {\bf m}\right\rangle$. Here and below we use the notation
$\left| A\right\rangle$
to denote states in the Fock subspace corresponding to the
unspecified ``other states" in
the single-particle basis. There are $2^{N}$ components in the
superposition (\ref{cis}), one for
every value of ${\bf x}.$ The states $\left| A_{{\bf x}}\right\rangle $
should be
determined by the requirement that the ground states are annihilated by
all
positive-energy annihilation operators $\Gamma _{E}$. Although we do
not know the
complete set of operators $ \Gamma_{E}$, we can now show that
\begin{equation}
\left\langle A_{\bf x}|A_{{\bf x}^{\prime }}\right\rangle =
{\frac{1}{2^{N} }}\,\delta _{{\bf x}{\bf x}^{\prime
}}.\label{axorthogonal}
\end{equation}

To see that, we first note that since the operators $X_{j}$ are
composed of finite-energy quasiparticle operators only, the matrix
element for any odd number of $X$ operators between any two ground
states must vanish. Thus,
\begin{equation}
\left\langle {\rm gs}_{\alpha }\left| X_{j_{1}}\right| {\rm gs}_
{\beta } \right\rangle
=\left\langle {\rm gs}_{\alpha }\left| X_{j_{1}}X_{j_{2}}X_{j_{3}}
\right| {\rm gs}_{\beta
}\right\rangle =\ldots=0 \label{oddxme}
\end{equation}
for arbitrary indices $\alpha ,\beta ,j_{1},\ldots $ Second, the matrix
elements of a
product of two {\it different} operators $X_{j}$ between states in the
ground-state
manifold are
\begin{eqnarray}
\left\langle {\rm gs}_{\alpha }\left| X_{j_{1}}X_{j_{2}}\right| {\rm gs}
_{\beta
}\right\rangle &=&\delta _{\alpha \beta
}\sum_{E>0}C_{E}^{j_{1}}C_{E}^{j_{2}\ast }  \nonumber \\
&=&2\delta_{\alpha\beta}\int \mathrm{d}{\bf r}\mathrm{d}{\bf r}^
{\prime }w_{j_{1}}^{(0)}({\bf
r})[w_{j_{2}}^{(0)}( {\bf r}^{\prime })]^*e^{\frac{i}{2}\left( \Omega _
{j_{1}}-\Omega
_{j_{2}}\right) }\sum_{E>0}u^\ast_{E}({\bf r})u_{E}({\bf r}^{\prime }).
\label{evenxme}
\end{eqnarray}
For sufficiently high energies the functions $u_{E}({\bf r})$ are
approximately plane
waves so that $\sum_{E}u^\ast_{E}({\bf r})u_{E}({\bf r}^{\prime })$ is
a short-ranged
function, presumably decaying exponentially with $\left| {\bf r}-{\bf r}
^{\prime }\right|
$, even for non-uniform superconductors. Then, for well-separated
vortices $j_1$ and
$j_2$, the matrix elements in Eq.\ (\ref {evenxme}) approximately
vanish. Put in
different words, the operation of the operators $X_{i}$ is spatially
localized around
vortex $i$, and thus two such operators operating near two distant
vortices generate
orthogonal excitations. Similarly, the matrix element of any other even
number of
different $X$ operators, taken with respect to any two ground states,
vanishes as well.
In Sec.\ \ref{nearcore-section}, we will also give a direct algebraic
proof of this
result which relies on a relation of the $X_i$ to operators
annihilating the ground
states.

The conditions in Eqs.\ (\ref{oddxme}) and (\ref{evenxme}) imply some
general conclusions
regarding the states $\left| A_{\bf x}\right\rangle$, which we first present for
the case of two vortices. There are four states $\left| \tau
\right\rangle$, labelled by
$\left| 00\right\rangle ,\left| 01\right\rangle ,\left| 10
\right\rangle ,\left|
11\right\rangle$.\cite{foot3} The phases encoding the vortex positions
are related by
$\Omega_{2}=\Omega_{1}+\pi$. The two ground states then take the
explicit form
\begin{eqnarray}
|m=0\rangle&=&\left( \left| 00\right\rangle - e^{-i\Omega _{1}}\left| 11
\right\rangle
\right) \left| A_{0}\right\rangle + e^{-i\Omega _{1}/2}\left( \left| 10
\right\rangle -
\left| 01\right\rangle \right) \left| A_{1}\right\rangle
\\
|m=1\rangle&=&\left( \left|00\right\rangle + e^{-i\Omega _{1}}\left| 11
\right\rangle
\right) \left| A_{1}\right\rangle + e^{-i\Omega _{1}/2}\left( \left| 10
\right\rangle +
\left| 01\right\rangle \right) \left| A_{0}\right\rangle.
\label{twovorwf}
\end{eqnarray}
The condition $\left\langle {\rm gs}_{\alpha }\left| X_{i}\right| {\rm
    gs} _{\alpha }\right\rangle =0$ implies that $\left\langle A_{0}|
  A_{1}\right\rangle =0$. In addition, $\left\langle {\rm gs}_{\alpha
  }\left| X_{1}X_{2}\right| {\rm gs}_ {\alpha }\right\rangle =0$
imposes $\left\langle A_{0}|A_{0}\right\rangle =\left\langle A_{1}
  |A_{1}\right\rangle =1/2$. Thus, while we cannot find the complete
wave functions of the ground states without a full solution of the BdG
equations, our procedure leads to the conclusion that the two ground
states are incoherent superpositions of the states $\left( \left|
    00\right\rangle \pm e^{-i\Omega _{1}}\left| 11\right\rangle
\right) $ with the states $e^{-i\Omega _{1}/2}\left( \left|
    10\right\rangle \pm \left| 01\right\rangle \right) $, with equal
weights to both components. There is an equal probability $1/4$ for
all four possible charge arrangements $\left| 00\right\rangle ,\left|
  01 \right\rangle ,\left| 10\right\rangle ,\left| 11\right\rangle $.
The parity of the particle number differs between the two basis
vectors of the ground-state manifold. However, this difference in
parity does not originate from the occupations of the two vortex-core
states. A local measurement of one of the two vortex-core states
cannot distinguish between the two ground states.

Generalizing to $2N$ vortices, it is easy to see that the requirement
that the expectation value of products of $X$ operators vanishes leads
to Eq.\ (\ref{axorthogonal}). Thus, all basis functions of the
ground-state manifold are an incoherent superposition of $2^{N}$
terms, of equal weight. Each of these terms is, by itself, a coherent
superposition of $2^{2N-1}$ possible occupations of the core states,
which constitute all possible occupations of a given parity. This
observation clarifies why there are $2^{N}$ ground states, rather than
$2^{2N}$. Within the Fock space of the core states there are $2^{N}$
eigenvectors for each eigenvalue ${\bf m} $, but a ground state is
generated by their incoherent superposition, {\it mixing all of them
  with equal weights.}

The wave functions in Eq.\ (\ref{cis}) give an explicit description of
the occupations of the core states. The operator
$\hat{N}(w_{i}^{(0)})\equiv (1/2)\left( \gamma_{i}+iX_i\right)
\left(\gamma_{i}-iX_i\right) $ counts the number of particles in the
$i$th core state. It is easy to see that
\begin{equation}
\left\langle {\bf m}^{\prime }\right| \hat{N}(w_{i}^{(0)})\left| {\bf
m} \right\rangle
=\left\langle {\bf m}^{\prime }\right| \left[ \hat{N} (w_{i}^{(0)})
\right] ^{2}\left|
{\bf m}\right\rangle =\frac{1}{2}\,\delta _{ {\bf m}{\bf m}^{\prime }}.
\end{equation}
Thus, in all possible ground states (including arbitrary superpositions
of the states
$\left| {\bf m}\right\rangle $) the occupation of each core state is
particle-hole
symmetric, with a probability of one half for being empty or occupied.

\section{Geometric phases and quantum entanglement in the evolution of
core states under vortex braiding
\label{entanglement-section}}

When vortices are adiabatically moved along closed trajectories,
ending with braiding of their positions, the ground state may
evolve in time away from the initial state, and the final ground
state may thus be different from the initial one. There are, in
principle, two contributions to this transformation of the ground
state.\cite{GurarieNayak} The first originates from its explicit
{\it multiply-valued} dependence on the phases $\Omega_i$ (explicit
monodromy). The second is through a nonabelian geometric vector
potential that gives rise to a nonabelian Berry phase. As is
always the case with geometric phases, only the sum of the
explicit monodromy and the Berry phase is observable and one can
split this sum between both contributions in any way desired by
choice of appropriate phase factors. We now show that the phase
choice we made in Eq.\ (\ref{zeromodespinor}) makes the second
contribution vanish, and proceed to calculate the first
contribution.

We need to prove that the geometric vector potential\cite{Wilczek2}
\begin{equation}
{\rm Im}\left\langle {\bf m}\left| \nabla _{{\bf R}_{i}}\right|
{\bf m}^\prime\right\rangle  \label{berry}
\end{equation}
vanishes for all ${\bf m},{\bf m}^\prime, {\bf R}_i$. The states $\left
|{\bf
m}\right\rangle$ depend on ${\bf R}_i$ through the phases $\Omega_i$
and through the
functions $w_i^{(0)}({\bf r})$. Since the $w_i^{(0)}({\bf r})$ are real
(up to some
trivial global phase factor), their derivative does not contribute in
Eq.\ (\ref{berry}),
and we may write $\nabla _{{\bf R}_{i}}=\sum_{j} (\nabla _{{\bf R} _{i}}
\Omega _{j})\
\frac{\partial }{\partial \Omega _{j}}$ and compute the matrix element
\begin{eqnarray}
    \langle {\bf m}| \frac{\partial }{\partial \Omega _{j}}| {\bf
       m}^\prime\rangle &=& \delta_{{\bf m}{\bf m}^\prime}\sum_{\bf x}
       \langle A_{\bf x}|\frac{\partial}{\partial \Omega _{j}}| A_{{\bf
x}}
       \rangle + {1\over 2^N}\sum_{\bf x}\langle {\bf m},{\bf x}|\frac
{\partial}
       {\partial \Omega _{j}}|{\bf m}^\prime ,{\bf x}\rangle.
\end{eqnarray}
The first term on the right-hand side is diagonal in ${\bf m},{\bf
  m'}$ and otherwise independent of ${\bf m}$. It therefore leads only
to abelian phase factors. Using the explicit states in Eq.\
(\ref{cxstate}) one finds that also the diagonal elements ${\bf
  m}={\bf m}^\prime$ of the second term are independent of ${\bf m}$.
The only contribution to the nonabelian part of the Berry phase can
therefore arise due to the off-diagonal contributions ${\bf m}\neq{\bf
  m}^\prime$ of the second term. However, these vanish as one may
verify by using the explicit states in Eq.\ (\ref{cxstate}).  Thus,
when the wave functions are written as in Eqs.\ (\ref{cxstate}) and
(\ref{cis}) the only phase factors that lead to non-trivial unitary
transformations arise from the explicit dependence of these wave
functions on the phases $\Omega_i$.

More explicitly, Eq.\ (\ref{cxstate}) shows that {\it the part of the
  wave function in which the core state $i$ is occupied has an
  amplitude proportional to $e^{-i\Omega_i/2}$, while the part in
  which this state is empty does not depend on $\Omega_i$. Since the
  ground states involve superpositions of empty and occupied core
  states, when the phase $\Omega_i$ accumulates a $2\pi$ shift, a
  relative minus sign is introduced between the components of the
  superpositions in which core state $i$ is empty and occupied, and
  the ground state does not necessarily come back to itself.} The
source of the evolution from one ground state to another is then in
the phases between different components of the wave function, and not
in any change of occupation of states.

It is instructive to examine in detail the case of the $i$th vortex
encircling the $j$th vortex, for which
\begin{eqnarray}
\Omega_{i}&\rightarrow&\Omega_{i}+2\pi \nonumber \\
\Omega_{j}&\rightarrow&\Omega_{j}+2\pi. \label{phasechangewinding}
\end{eqnarray}
This change of phase affects both the states $\left |{\bf m},{\bf
x}\right \rangle $ and the states $\left |A_{\bf x}\right\rangle$,
according to
\begin{eqnarray}
\left |{\bf m},{\bf x}\right
\rangle&\rightarrow&(2i\gamma_{i}X_{i})(2i\gamma_{j}X_{j})\left
|{\bf m},{\bf x}\right \rangle\label{phase-changes-m}\\
\left |A_{\bf x}\right\rangle&\rightarrow&{{\hat{\cal B}}_{ij}}\left |A_
{\bf x}\right\rangle\label{phase-changes-a}
\end{eqnarray}
where ${{\hat{\cal B}}_{ij}}$ is an operator that acts on the
``other", non-core, states, only. The final state $\sum_{\bf
  x}(2i\gamma_{i}X_{i})(2i\gamma_{j}X_ {j})\left |{\bf m},{\bf
    x}\right \rangle{{\hat{\cal B}}_{ij}}\left |A_{\bf x}
\right\rangle$ must be a ground state. However, the operators
$X_i,X_j$ generate excitations above the ground states. The operator
${\hat{\cal B}}_{ij}$ must then be an operator that annihilates these
excitations. More precisely, since the states $|A_{\bf x}\rangle$ do
not depend on ${\bf m}$, the operator $X_iX_j{\hat{\cal B}}_{ij}$ must
be a $c$-number within the subspace of ground states. In fact, for
the norm of the ground state to be conserved during the braiding of
vortices, the magnitude of this $c$-number must be unity, i.e.,
\begin{equation}
2X_iX_j{\hat{\cal B}}_{ij}=e^{i\phi_a}\label{xarelationw}
\end{equation}
with $\phi_a$ being an abelian phase. In this way, we recover the
known unitary transformation for winding of two vortices,
$2\gamma_i\gamma_j$ given in Eq.\ (\ref{unit-trans}) [see also App.\
\ref{unitary}]. This shows that despite the appearance of Eq.\
(\ref{unit-trans}), this transformation does not involve any changes
of core-state occupations, but only changes in relative phases between
the components of a superposition, each of which has different
core-state occupations.

In the subsequent sections, we will study the states $|A_{\bf x}
\rangle$ much more explicitly by introducing the ``near-core"
single-particle states. We will find that these states are
structurally very similar to the ground states $|{\bf m}\rangle$ and
this allows us to give explicit expressions for the operators ${\hat
  {\cal B}}_{ij}$ (up to an abelian phase factor). These expressions
do indeed satisfy Eq.\ (\ref {xarelationw}).

The above observations allow us to conclude that the ground states spanned
by the basis in Eq.\ (\ref{cis}) are states {\it in which the
  occupation of the core states at different vortices are fully
  entangled}, in the following sense: There is no ground state of the
system in which the parity of the number of particles is well defined,
and in which the occupations of two subsets of core states are
disentangled from one another. If such a state were to exist, its wave
function could be written as
\begin{equation}
\hat{C}_{1}\hat{C}_{2}\hat{C}_3\left| {\rm vac}\right\rangle, \label
{disentangled}
\end{equation}
where $\hat{C}_{1}$ is an operator that acts only on the core states
belonging to the first subset, $\hat{C}_{2}$ is an operator that acts
only on core states belonging to the second subset, and $\hat{C}_3$ is
an operator that acts solely on the remaining states (core states or
other states) not included in any of the two subsets. However, for the
state in Eq.\ (\ref {disentangled}) to have well-defined
particle-number parity, the states created by $\hat{C}_{1}$ or
$\hat{C}_{2}$ must each have definite parity. By considering the
effect of an encircling in which one vortex (say, the $i$th vortex)
from the first subset winds around another vortex of the second subset (say,
the $j$th vortex), we find that this cannot happen. The unitary
transformation $2\gamma _ {i}\gamma _{j}$ corresponding to this
transformation changes the parity of the particle number for both the
first and the second subset, while we showed that the transformation
of the ground state is a consequence of changes in phases only, rather
than changes in core states occupation. Consequently, a state of the
form in Eq.\ (\ref {disentangled}) cannot be a ground state.

In some sense, the entanglement of the occupations of core states is
signaled by Eq.\ (\ref{occupationofalpha}). As seen from that
equation, the ground states $\gamma_{2j-1}\left |{\bf m}\right\rangle$
and $\gamma_{2j}\left |{\bf m}\right\rangle$ differ from one another
only by a phase factor, despite the fact that they are obtained from
the ground state $\left |{\bf m}\right\rangle$ by the application of
two Majorana operators localized very far from one another.

Interestingly, despite the entanglement, the occupations of different
core states $i$ and $j$ are
uncorrelated,
\begin{equation}
\langle {\rm gs} |\hat {N}(w_i^{(0)})\hat {N}(w_j^{(0)})\left| {\rm gs}
\right\rangle=
\langle {\rm gs} |\hat {N}(w_i^{(0)})\left| {\rm gs}\right\rangle
\langle {\rm gs} |\hat
{N}(w_j^{(0)})\left| {\rm gs}\right\rangle=\frac{1}{4}. \label{no-
correlation}
\end{equation}
This lack of correlations persists also to higher-order correlators.

Furthermore, we can conclude that the total number of particles in the
core states, counted by the operator ${\hat N}_{\rm core}=
\sum_{i=1}^{2N}{\hat N} (w_i^{(0)})$, does not have a well-defined
parity. Rather, it has equal probabilities for being even and odd,
irrespective of what is the parity of the total number of particles in
the ground state. To see that, we consider the particle-number parity
operator $\exp(i\pi{\hat N}_{\rm core})$. Due to the lack of
correlations between different vortices,
\begin{equation}
\langle {\rm gs} |\exp{(i\pi{\hat N}_{\rm core})}   | {\rm gs}
\rangle=\prod_{j=1}^{2N}
\langle {\rm gs} |\exp{[i\pi{\hat N}(w_i^{(0)})]}| {\rm gs}\rangle=0.
\end{equation}
Thus, the parity of the ground states cannot be determined by a
measurement of the occupation of the core states alone.

Despite the inherent quantum entanglement of the ground states, it
appears impossible to formulate corresponding Bell inequalities.
Each core state defines a two-dimensional Hilbert space, similar
to a spin-$\frac{1}{2}$. However, the operators associated with
these spaces at different cores, the {\it fermionic} operators
$\gamma_j$ and $X_j$, do not commute. Thus, their classical analogs
are ill-defined.\cite{Aharonov-vaidman}

What happens when two vortices are interchanged can be analyzed along
similar lines to the case of one vortex encircling another. We now
analyze the case of interchanging vortices from the same pair, $2i-1$
and $2i$. Note that choosing the vortices from the same pair does not
imply any loss of generality, both because for any interchange we can
choose a pairing such that the two vortices are from the same pair, and
because our considerations will eventually lead to an operator
expression which is independent of the particular choice of pairing.

As the adiabatic motion of vortices does not involve any tunneling of
particles between vortex cores, an interchange of the position of
vortices interchanges the occupation of their core states. 
When both core states are occupied, the interchange is
accompanied by a factor of $(-1)$, since two fermions interchange
positions. In addition, one of the vortices necessarily crosses the
cut line of the phase of the other, changing this phase by $2\pi$
(cf.\ App.\ \ref{unitary}). The phase of the other vortex remains
intact. Implementing these transformations in Eq.\ (\ref{cxstate}),
the term $[ 1+i\,(-1)^{x_{2j}} e^{-i(\Omega _{2j-1}+\Omega _{2j})/2
}c_{2j-1}^{\dagger}c_{2j}^{\dagger}]$ remains unaffected by the
interchange, while the relative sign between the two components
changes in the term $[ e^ {-i\Omega
  _{2j-1}/2}c_{2j-1}^{\dagger}+i\,(-1)^{x_{2j}}e^{-i\Omega_{2j}/2}c_{2j}^
{\dagger}]$.

These transformations of the core-state wave functions are implemented
by the operator $e^{\pi\gamma_{2j-1}\gamma_{2j}/2}e^{\pi
  X_{2j-1}X_{2j}/2}$, acting on Eq.\ (\ref{cxstate}). By contrast, as
found in Ref.\ (\onlinecite{Ivanov}) (reviewed in App.\
\ref{unitary}), the unitary transformation that enacts the vortex
interchange on the states in Eq.\ (\ref{cis}) is
$e^{\pi\gamma_{2j-1}\gamma_{2j}/2}$. As in the case of encircling, the
difference between these two transformations is an operator $\hat{\cal
  B}$ that acts on the $\left | A_{\bf x}\right \rangle$ part of the
wave function in Eq.\ (\ref{cis}).

The states $\left| A_{\bf x}\right\rangle$ together with the
operators $\hat{\cal B}$ that act on them when vortices wind and
interchange are the subject of the following sections. We define a
set of single-particle states adjacent to the core state for each
vortex (``near-core states"), and show that the operator
$\hat{\cal B}$ affects the occupation of these states in exactly
the same way as the operators $2\gamma_{i}\gamma_{j}$ and
$e^{\pi\gamma_{2j-1}\gamma_{2j}/2}$ affect the occupation of the
core states for vortex winding and encircling, respectively. In
fact, we find the corresponding operators $\hat {\cal B}$ to have
the same functional forms, but with the Majorana operators
$\gamma_{i}$ and $\gamma_{j}$ replaced by analogous Majorana
operators associated with the near-core states. We explain this in
terms of a ``self-similar" structure of the ground-state
wave functions.

\section{States near the core\label{nearcore-section}}

Our information on the states $\left| A_{\bf x} \right\rangle$
introduced in Eq.\ (\ref{cis}) has so far relied on the conditions in
Eqs.\ (\ref{oddxme}) and (\ref{evenxme}) for matrix elements of
products of the operators $X_{i} $. These operators were constructed
to have two main virtues: (i) They create and annihilate particles in
the core states only. (ii) They create and annihilate only excitations
(Bogolubov quasiparticles) with $E>0.$ We now take these operators as
a starting point in a scheme by which we define the set of mutually
orthogonal single-particle states $w_{i}^{(k)}({\bf r})$, where the
subscript $i$ refers to the vortex number and the superscript $k$
enumerates the iterations in our scheme. The new single-particle
states remain localized near the vortex cores and provide insight into
the nature of the states $\left| A_{\bf x }\right\rangle$.

We get to these states by using the $X_i$'s to construct a set of
operators $Y_i^{(k)}$ that, unlike the Majorana $X_i$'s,
annihilate the ground states,
\begin{equation}
Y_{i}^{(k)}\left| {\bf m}\right\rangle =0
\label{allysannihilatem}
\end{equation}
for all $i,k,{\bf m}$. Furthermore, these operators anti-commute $\{Y_
{i}^{(k)},Y_{i^{ \prime }}^{(k^{\prime })}\}=0$ and each operator
$Y_i^{(k)}$ creates and annihilates particles only in the two basis
states $w_{i}^{(k-1)}({\bf r})$ and $w_ {i}^{(k)}({\bf r})$. By virtue
of the conditions in Eq.\ (\ref{allysannihilatem}), these operators
specify the ground-state wave functions in the subspace spanned by the
states $w_{i}^{(k)}({\bf r})$.

Our scheme starts from the expansion of the operators $X_{i}$ in Eq.\
(\ref{xdecomposed}). We define the two sets of operators
\begin{eqnarray}
Y_{i}^{(1)} &=&\sqrt{2}\sum_{E>0}C_{E}^{i}\Gamma _{E}  \label{Ydef} \\
Z_{i}^{(1)} &=&i\sum_{E>0}\,\left( C_{E}^{i}\Gamma
_{E}-C_{E}^{i\ast }\Gamma _{E}^\dagger\right).\label{Zdef}
\end{eqnarray}
The operators $Y^{(1)}_{i}$ annihilate all the ground states since
they are constructed from (positive-energy) annihilation operators
only. The operators $Z^{(1)}_{i}$ are, by construction, Majorana operators,
and can thus be written as
\begin{equation}
Z_{i}^{(1)}={\frac{1}{\sqrt{2}}}\int \mathrm{d}{\bf r}\left[ w_{i}^{(1)}
({\bf r})\psi ({\bf r})+{\rm
h.c.}\right].  \label{defw1}
\end{equation}
This defines a set of $2N$ single-particle states
$w_{i}^{(1)}({\bf r})$. We prove in appendix \ref{orthogonal} that for well separated vortices
the $w_{i}^{(1)}({\bf r})$ are mutually orthogonal as well as
orthogonal to the core states $w_{i}^{(0)}({\bf r})$. Thus, we can extend our single-particle
basis by adding to it the $2N$ states $w_{i}^{(1)}({\bf r })$.
Note that for conciseness of notation, we absorb the phase factors
$\exp\left(i\Omega_i/2\right)$ into the definition of the states
$w_i^{(k)}({\bf r})$ throughout this section. We comment below on
how they should be reinstated.

Since $X_i=(1/\sqrt{2})[Y_i^{(1)}+Y_i^{(1)\dagger}]$, the condition
\begin{equation}
Y_{i}^{(1)}\left| {\rm gs}_{\alpha }\right\rangle =0  \label
{yannihilatem}
\end{equation}
implies immediately that any combination of different operators $X_{i}$ has zero
matrix elements
between ground states as required in Sec.\ \ref{core-section}. Using
the relation
$Y_{i}^{(1)}=(1/\sqrt{2})(X_{i}-iZ_{i}^{(1)})$, we can write
\begin{eqnarray}
Y_{i}^{(1)} =\frac{i}{2}\int \mathrm{d}{\bf r}\left\{ \left(
w_{i}^{(0)}({\bf r})-w_{i}^{(1)}( {\bf r})\right) \psi({\bf r})
-\left( [w_{i}^{{(0)}}({\bf r} )]^\ast+[w_{i}^{{(1)} }({\bf
r})]^\ast\right) \psi ^\dagger({\bf
r})\right\}.\label{Ydefadvanced}
\end{eqnarray}
Thus, these $2N$ operators affect the occupations of the states
$w_{i}^ {(0)}({\bf r })$ and $w_{i}^{(1)}({\bf r})$ only.

To iterate this process it is helpful to summarize the steps leading
to the definition of $w_{i}^{(1)}({\bf r})$ through $Y_{i}^{(1)}$,
using the concise spinor representation for the operators.
We start with the operators $\gamma _ {i}={\frac{1}{ \sqrt{2}}}(
  w_{i}^{(0)}({\bf r}), [ w_{i}^{(0)}({\bf r}) ] ^{\ast }
)$ in Eq.\ (\ref{gamma}) which act on the occupation of the
$w_ {i}^{(0)}({\bf r})$ only. In Eq.\ (\ref{xoperators}) we define the
operators $X_{i} =i\sigma _{z}\gamma
_{i}={\frac{1}{\sqrt{2}}}(iw_{i}^{(0)}({\bf r}), -i [w_{i}^
    {(0)}({\bf r})] ^{\ast })$ acting on the occupations
of the same single-particle states ($\sigma_z$ is a Pauli
matrix). By construction, the spinors corresponding to $X_{i}$ are
orthogonal to those corresponding to $\gamma _{i}$, implying that the
corresponding operators anti-commute. Then, in Eq.\ (\ref{Ydef}), we
extract from the operators $X_{i}$ the parts $Y_{i}^{(1)}$ that
annihilate the ground state. The operators $Y_{i}$, in turn, are
written in Eqs.\ (\ref{Zdef})-(\ref {Ydefadvanced}) as sums of
hermitian operators $X_{i}$ and anti-hermitian operators $iZ_{i}^{(1)}
$, and finally, we defined $ w_{i}^{(1)}({\bf r})$ through the
Majorana operators $Z_{i}^ {(1)}.$ In spinor representation, these
last three steps can be recast as $ Z_{i}^{(1)}=- S\sigma _{z}\gamma
_{i}$ with
\begin{equation}
S=\sum_{E\ne 0}{\rm sgn}E\,\left| {u_{E}\atop v_{E}}\right\rangle
\left\langle {u_{E} \atop v_{E}}\right|. \label{opA}
\end{equation}
The operator $S$ is a difference of two projection operators. The positive (negative) energy part of the  sum projects to the subspace of positive (negative) energy BdG solutions.

We now iterate this process to define states $w_{i}^{(k)}({\bf r})$
and a set of operators $Y_{i}^{(k)}$ for which $Y_{i}^{(k)}\left| {\rm
    gs}_{\alpha } \right\rangle =0$.  This is achieved by generating
the set of Majorana operators
\begin{equation}
Z_{i}^{(k)}={\frac{1}{\sqrt{2}}}\left(
\begin{array}{c}
w_{i}^{(k)}({\bf r}) \\
\left[ w_{i}^{(k)}({\bf r})\right] ^{\ast }
\end{array}
\right) =\left[ -S\sigma _{z}\right] ^{k}\gamma _{i}. \label{zk}
\end{equation}
When writing out $S$ explicitly in real space representation, according to Eq.\ (\ref{opA}), it
contains energy sums of the type that has been discussed following
Eq.\ (\ref{evenxme}). By the same arguments employed there, we can
conclude that as a
function of the two coordinates ${\bf r}$ and ${\bf r'}$, the operator $S$
is short ranged, i.e., decays fast for large $|{\bf r}-{\bf r'}|$.
Thus, $w_{i}^{(k)}({\bf r})$ is also localized around the $i$th
vortex although its distance from the vortex core increases with $k$.
Since the construction assumes that states $w_i^{(k)}({\bf r})$
localized around different vortices do not overlap, there exists an
upper limit to the number of iterations. We denote the last iteration number by $L$.

The functions $w_{i}^{(k)}({\bf r})$ are studied in more detail in
App.\ \ref {orthogonal}. We find that for iterations $k<L$, the various
$w_{i}^{(k)}( {\bf r})$ are mutually orthogonal,
\begin{equation}
\langle w_{i}^{(k)}|w_{j}^{(k^{\prime })}\rangle =\delta _{ij}\delta _
{kk^{\prime }}.
\label{wortho}
\end{equation}
As long as the spatial extent of the states $w_i^{(k)}({\bf r})$ is small compared to the distance between vortices, the dependence of the near-core states $w_i^{(k)}({\bf r})$ on the
phases $\Omega_i$ is analogous to that of the core states.
Specifically, these phases can be made explicit by the replacement
\begin{equation}
       w_i^{(k)}({\bf r}) \to w_i^{(k)}({\bf r}) e^{i\Omega_i/2}.
\end{equation}
This can be seen from Eqs.\ (\ref{opA}) and (\ref{zk}) in combination
with the observation that for ${\bf r}$ in the vicinity of vortex $i$,
all finite-energy BdG solutions $u_E({\bf r})$ and $v_E({\bf r})$
depend on the vortex positions through the phase factor
$e^{i\Omega_i/2}$.

We also define
\begin{eqnarray}
X_i^{(k)} ={\frac{1}{\sqrt{2}}}\left(
\begin{array}{c}
iw_{i}^{(k)}({\bf r}) \\
-i\left[ w_{i}^{(k)}({\bf r})\right] ^{\ast }
\end{array}
\right)
\end{eqnarray}
and the operators
\begin{eqnarray}
Y_{i}^{(k)} &=& {\frac{1}{\sqrt{2}}}[X_{i}^{(k-1)}-iZ_{i}^{(k)}]
\nonumber
\\
&=&{\frac{i}{\sqrt{2}}}(1 + S) \sigma_{z}Z_{i}^{(k-1)}  \nonumber \\
&=&\frac{i}{2}\left[ \left( c_{i}^{(k-1)}-c_{i}^{(k)}\right) -\left(
c_{i}^{{(k-1)}\dagger }+c_{i}^{{(k)}\dagger }\right)\right] \label
{yoperator}
\end{eqnarray}
which annihilate the ground states [cf.\ Eq.\ (\ref{allysannihilatem})]. Here, the operators $c_i^{(k)}$
annihilate particles in the states $w_i^{(k)}({\bf r})$. This
iterative construction enlarges our single-particle basis. Starting
with the $2N$ core states $w_{i}^{(0)}({\bf r})$, we defined an
additional $L-1$ ``near-core'' states for every vortex.

The occupations of the newly-defined $L-1$ ``near-core" states are all
particle-hole symmetric. This can be seen by noting that the operator
$\hat{N%
}(w_{i}^{(k)})= (1/2)\left( Z_{i}^{(k)}+iX_{i}^{(k)}\right) \left(
Z_{i}^{(k)}-iX_{i}^{(k)}\right)$ counts the number of particles in the
state
$w_{i}^{(k)}({\bf r})$. Since
\begin{eqnarray}
\hat{N}(w_{i}^{(k)})&=&{\frac{1}{2}}+iX_{i}^{(k)}Z_{i}^{(k)}  \nonumber
\\
&=&{\frac{1}{2}}\left[1+ \left(
Y_{i}^{(k+1)\dagger}+Y_{i}^{(k+1)}\right)\left( Y_{i}^{(k)\dagger}-Y_{i}
^{(k)}\right) \right]
\end{eqnarray}
and since the operators $Y_{i}^{(k)\dagger}$ and $Y_{i}^{(k+1)}$
correspond
to orthogonal excitations, one finds
\begin{equation}
\left\langle {\bf m}^{\prime }\right| \hat{N}(w_{i}^{(k)})\left| {\bf
m} \right\rangle = \left\langle {\bf m}^{\prime }\right| \left[\hat{N}
(w_{i}^{(k)})\right]^{2}\left| {\bf m} \right\rangle
=\frac{1}{2}\,\delta_{{\bf m}{\bf m}^{\prime }}.
\end{equation}

In the next section we solve Eq.\ (\ref{allysannihilatem}) to extract
the ground-state
wave functions in the Fock subspace spanned by the near-core states
$w_i^{(k)}({\bf r})$.

\section{Ground-state wave functions and nonabelian statistics}
\label{near-core-wf-section}

Solving Eq.\ (\ref{allysannihilatem}), we now show that the structure
we found in Sec.\
\ref{core-section} for the occupations of the core states repeats
itself for the
near-core states. This allows us to complete the argument that the
effect of vortex
braiding can be understood explicitly in terms of the geometric phases
$\Omega_i$.

The detailed form of the ground-state wave function depends on the
convention for the
order in which creation operators of various states act on the vacuum.
Eq.\
(\ref{cis}) positions all creation operators for the core states to the
left of creation
operators for other states, and Eq.\ (\ref{labelmx}) specifies the
order in which core
states are being filled. In that spirit, the creation operators for the
first iteration
of near-core states $w_i^{(1)}({\bf r})$ are going to be positioned to
the right of those
of the core states, and so on with increasing number of the iteration.
Within each
iteration, we follow the convention defined in Eq.\ (\ref{labelmx}) for
the core states.

We first determine the occupations of the near-core states of the first
iteration,
$w_i^{(1)}({\bf r})$ in the states $|A_{\bf x}\rangle$. To this end, we
define creation
and annihilation operators from pairs of the operators $Z_i^{(1)}$ and
$X_i^{(1)}$,
\begin{eqnarray}
     \delta_{2j} &=& {1\over \sqrt{2}} (Z_{2j-1}^{(1)} - i Z_{2j}^{(1)})
\\
     \eta_{2j}&=& {1\over \sqrt{2}} (-i X_{2j-1}^{(1)} +  X_{2j}^
{(1)}).
\end{eqnarray}
Then we can express the states $|A_{\bf x}\rangle$ in terms of the
states
\begin{equation}
    |{\bf m},{\bf x}\rangle^{(1)} = (\delta_{2N}^\dagger)^{m_{2N}}\cdots
(\delta_{4}^\dagger)^{m_4}
   (\delta_{2}^\dagger)^{m_2} (\eta_{2}^\dagger)^{x_{2}}(\eta_{4}
^\dagger)^{x_4}\cdots (\eta_{2N}^\dagger)^{x_{2N}} |{\bf m}=0,{\bf x}=0
\rangle^{(1)},
\end{equation}
where the superscript on the state indexes the iteration. The
conditions
$(Y^{(1)}_{2j-1}\pm i Y^{(1)}_{2j})|{\bf m}\rangle =0$ for the ground
states can be
rewritten as [cf.\ Eq.\ (\ref{yoperator})]
\begin{eqnarray}
     (\beta_{2j} - \delta_{2j}^\dagger)|{\bf m}\rangle &=& 0 \\
     (\beta^\dagger_{2j} + \delta_{2j})|{\bf m}\rangle &=& 0.
\label{cond-y}
\end{eqnarray}
Applying these conditions to the ground states, we find
\begin{equation}
|A_{\bf x}\rangle = \sum_{{\bf x}^\prime}|{\bf x},{\bf x}^\prime
\rangle^{(1)} |A_{{\bf x}^\prime}\rangle^{(1)},\label{axexplicit}
\end{equation}
in terms of further states $|A_{{\bf x}^\prime}\rangle^{(1)}$
which contain the occupations of the remaining near-core states as
well as the other states. As expected, Eq.\ (\ref{axexplicit})
satisfies Eq.\ (\ref{axorthogonal}). Iterating this procedure, we
get
\begin{equation}
|A_{\bf x}\rangle^{(1)} = \sum_{{\bf x}^\prime}|{\bf x},{\bf
x}^\prime \rangle^{(2)} |A_{{\bf
x}^\prime}\rangle^{(2)},\label{ax1explicit}
\end{equation}
and so on, as long as near-core states from different vortices do not
overlap.  Consequently, we can write the ground-state wave functions
as
\begin{equation}
   |{\bf m}\rangle = \sum_{{\bf x}^{(0)}} |{\bf m},{\bf x}^{(0)}\rangle^
{(0)}
              \sum_{{\bf x}^{(1)}} |{\bf x}^{(0)},{\bf x}^{(1)}\rangle^
{(1)}
              \sum_{{\bf x}^{(2)}} |{\bf x}^{(1)},{\bf x}^{(2)}\rangle^
{(2)} \cdots
              \sum_{{\bf x}^{(L)}} |{\bf x}^{(L-1)},{\bf x}^{(L)}
\rangle^{(L)}
              |A_{{\bf
              x}^{(L)}}\rangle^{(L)}.\label{self-similar-wf}
\end{equation}
This structure of the ground states is helpful for analyzing the
effect of vortex braiding.

As discussed in Sec.\ \ref{entanglement-section}, the effect of
encircling vortex $i$ by vortex $j$ on the ground states $|{\bf
m}\rangle$ is given by the unitary transformation
$2\gamma_i\gamma_j$  [Eq.\ (\ref{unit-trans})]. In Eqs.\
(\ref{phase-changes-m}) and (\ref{phase-changes-a}), we decomposed
this transformation into two factors, one affecting the core
states, and the other affecting the other states.

The ``self-similarity" of the wave function, evident in Eqs.\ (\ref
{axexplicit}) - (\ref{self-similar-wf}), suggests that the effect of
vortex encircling on the states $|A_{\bf x}\rangle$ is analogous to
its effect on the ground states $| {\bf m}\rangle$ themselves.
Clearly, the analog of the operators $\gamma_i$ in the first
generation of near-core states are the operators $Z_i^{(1)}$. Thus,
one expects
\begin{equation}
    \hat{\cal B}_{ij} = 2Z_i^{(1)}Z_j^{(1)}
\end{equation}
and the associated operator identity
\begin{equation}
    2\gamma_i\gamma_j = (2i\gamma_i X_i)(2i\gamma_j X_j)(2Z^{(1)}_iZ^
{(1)}_j)
\label{encirclingid}
\end{equation}
when applied to any ground state. Indeed, the proof of this identity
follows from the
fact that $Y_i=(1/\sqrt{2})(X_i-iZ_i^{(1)})$ annihilates any ground
states so that, when
acting within the subspace of ground states, one has $0= \sqrt{2} X_iY_i = 1/2 -iX_iZ_i^
{(1)}$ or
$2iX_iZ_i^{(1)}=1$.

An analogous picture emerges for the exchange of vortices. Here, we
found that the explicit effect of exchange on the states $|{\bf
  m},{\bf x}\rangle$ is equivalent to the action of the operator
$\exp\{\pi\gamma_i\gamma_j/2\}\exp\{\pi X_iX_j/2 \}$. On the other
hand, the effect of the interchange on the ground states is given by
the operator $\exp\{\pi\gamma_i\gamma_j/2\}$. Due to the
``self-similarity" of the wave function, the connection between these
two operators is expected to be furnished by
\begin{equation}
  \hat{\cal B} = \exp\{\pi Z^{(1)}_iZ^{(1)}_j/2\}.
\end{equation}
Indeed, there exists a corresponding operator identity
\begin{equation}
\exp\{\pi\gamma_i\gamma_j/2\}=\exp\{\pi\gamma_i\gamma_j/2\}\exp\{\pi
X_iX_j/2\}\exp\{\pi
Z^{(1)}_iZ^{(1)}_j/2\}
\end{equation}
valid when acting within the subspace of ground states. Its proof
uses the same ingredients as in the case of encircling.

For both vortex encircling and exchange, this procedure can now be
repeated for the near-core states obtained in higher iterations. In
this way, it follows that the effect of vortex encircling or exchange
on any core or near-core states is entirely contained in the geometric
phases $\Omega_i$.

This discussion clarifies the difference between unitary transformations composed of zero-energy Majorana operators, say,
$\gamma_{2j-1}\gamma_{2j}$, and unitary transformations composed of non-zero energy Majorana operators, such as $X_{2j-1}^{(k)}X_{2j}^{(k)}$ and $Z_{2j-1}^{(k)}Z_{2j}^{(k)}$. When acting on a ground state, the first one leaves the system within the subspace of ground states, while the last two excite the system. Any unitary transformation that may be implemented by means of an adiabatic vortex braiding leaves the system in the ground state subspace. That is the case for $\gamma_{2j-1}\gamma_{2j}$, whose effect on a ground state may be reduced to a series of phase changes associated with vortex braiding. By contrast, the effect of the other two operators may not be implemented by an adiabatic braiding of vortices.

The number of generations $L$  appearing in Eq. (\ref{self-similar-wf}) is chosen such that the states  $w_i^{(k)}({\bf r})$ generated by our iterative process do not become extended enough to overlap. If the iterative process is carried out further, states from different vortices start overlapping, and therefore cannot be used as basis states in a single-particle basis. In principle, these states may be orthogonalized by means of a Gram-Schmidt-type process. The states resulting from that process, however, would not necessarily share the properties of the states $w_i^{(k)}({\bf r})$ for $k<L$, namely their occupation may not be particle-hole symmetric, and the way they are affected by vortex braiding may be different.

\section{Summary}
\label{conclusions}

The composite-boson theory of the fractional quantum
Hall effect at filling fractions $\nu=1/m$ ($m$ odd) employs a
Chern-Simons transformation to map the electronic system to a
superfluid of composite bosons. It explains the fractional
statistics of the excitations of these states by describing them
as charged vortices in that superfluid, and the
statistical phase as the effect of the charge of one vortex on the
geometric phase accumulated by another vortex as it moves
adiabatically.

The Moore-Read theory describes the $\nu=5/2$ state also as a
superfluid. Again, charged excitations are vortices in this
superfluid, and these vortices accumulate geometric phases as they
move. However, in this case the effective bosons forming the superfluid are
Cooper pairs of composite fermions. As a consequence, this state has
excitation modes that involve the breaking of Cooper pairs into two
composite fermions. These modes are solutions of the Bogolubov-deGennes
equations. In the presence of well-separated vortices, each vortex
gives rise to one zero-energy solution. These zero-energy solutions
represent Majorana operators, operating on the core state of their
vortex. For $2N$ vortices, these zero modes lead to a $2^N$-fold
degeneracy of the ground state in the presence of vortices. The $2N$
core states define a $2^{2N}$-dimensional Fock space, since each core
state may be occupied or empty. We found that each ground state has an
equal probability for each of the $2^{2N}$ possibilities to occupy the
core states, and that the way these possibilities are superposed is
such that (at least for ground states with a definite parity of the
total particle number) the occupation of core states in one vortex is
entangled with the occupations of all core states in all other
vortices. We showed that it is the relative phases between the
amplitudes for different occupations that distinguishes between the
various ground states, and that it is these relative phases that vary
as the vortices move. This allows the system to transform between
ground states due to vortex braiding. It is interesting to observe that unlike the case of ordinary quantized Hall states, there seems to be no direct relation between the charge carried by the vortices and their non-abelian statistics.

While the zero-energy Majorana modes affect only the occupations of
the core states of the vortices, we showed that it is possible to
extract some general information regarding the occupations of other
single-particle states without explicitly solving the
Bogolubov-deGennes equations. In particular, we constructed a set of
single-particle states which are localized near the vortex cores and
whose occupations are also maximally uncertain (all possibilities
of their occupations are equally probable) and maximally entangled.
The definition of these near-core states allowed us to explicitly
relate our picture of nonabelian statistics, based on entanglement and
geometric phases, to previously existing approaches based on
representations of vortex braiding in terms of unitary transformations that create and annihilate particles in the core states. The elucidation of the role of quantum entanglement and geometric phases may be useful for understanding the effect of decoherence on non-abelian statistics, a question which is  of relevance in the context of topological quantum computation. \cite{Kitaev}

The Moore-Read state is only the simplest example of proposed FQHE
wave functions \cite{ReadRezayi} whose excitations satisfy nonabelian
statistics. As in this example, nonabelian statistics is generally
associated with a degeneracy of the ground state.  However, the other
proposed states lack a description in terms of second quantization,
and in the absence of such a description it is hard to generalize our
analysis for the Moore-Read state to the entire set of nonabelian
states. We believe, however, that a second-quantization formulation of
other nonabelian states would lead to a similar picture.

\begin{acknowledgments}
  We thank J.E.\ Avron, B.I.\ Halperin, and N.\ Read for instructive
  discussions. AS acknowledges support of the US-Israel Binational
  Science Foundation and the Israel Science Foundation. FvO thanks the
  Einstein and Submicron Centers at the Weizmann Institute for
  hospitality and support (LSF project HPRI-CT-2001-00114) on several
  occasions, including a Michael visiting professorship during
  completion of this work. He was also supported by the DFG
  Schwerpunkt ``Quanten-Hall-Systeme'' as well as the Junge Akademie.
  EM acknowledges support from the LSF project HPRI-CT-1999-69.
\end{acknowledgments}

\appendix

\section{Unitary transformations for vortex encircling and interchange}
\label{unitary}

In this appendix, we review for completeness how to construct the
unitary transformations $U$ associated with particle exchange and
encircling.\cite{Ivanov} For this purpose, we analyze the time
evolution of the many-particle state within the degenerate subspace of
ground states as the vortices adiabatically traverse trajectories that
start and end in the same set of positions. The unitary transformation
is defined by the relation between the final state $|\psi(t=T)\rangle$
and the initial state $|\psi(t=0) \rangle$,
\begin{equation}
    |\psi(t=T)\rangle = U |\psi(t=0)\rangle.
\end{equation}
Correspondingly, the time evolution of the operators $\gamma_i$
spanning the degenerate
subspace is given by
\begin{equation}
    \gamma_i(t=T)= U \gamma_i(t=0) U^\dagger.
\end{equation}
The time evolution of the operators $\gamma_i$ can be readily read off
from their
definition Eq.\ (\ref{gamma}) and the explicit form Eq.\ (\ref
{zeromodespinor}) of the
zero-energy spinors of the Bogolubov-deGennes equations. This can be
used to identify the
operators $U$ up to an abelian phase. We note that with our choice of phase for the zero-energy spinors in Eq.\
(\ref{zeromodespinor}), their Berry phase during adiabatic phase
evolution vanishes and
the phase evolution is given solely by the explicit monodromy.

Before constructing $U$ for exchange trajectories or encircling, we
first construct
unitary operators from the set of $\gamma_i$'s which are helpful in
giving explicit
expression for the operators $U$.
\begin{itemize}
\item  \bigskip The unitary transformation $u_k=\sqrt{2} \gamma_k$ adds
a minus sign to
all operators $\gamma_j$ with $j\neq k$. If $j=k$, it leaves the
operator unchanged:
\begin{equation}
u_{k}\gamma_j u_{k}^\dagger=\left\{
\begin{array}{cc}
\gamma_k & {\rm when }\,\,\,\,\,j=k \\
-\gamma_j & { \rm otherwise}
\end{array}
\right.
\end{equation}
Obviously $\left\{ u_{k},u_{k^{\prime }}\right\} = 2\delta _{k k^
{\prime }}$.

\item  The unitary transformation $u_{ij}=\gamma_i+\gamma_j$
interchanges $\gamma_i$ with
$\gamma_j$ and adds a minus sign to all other operators $\gamma_k$ with
$k\neq i,j$:
\begin{equation}
u_{ij}\gamma_k u_{ij}^\dagger=\left\{
\begin{array}{cc}
\gamma_i & {\rm when }\,\,\,\,\,\, k=j \\
\gamma_j & {\rm when }\,\,\,\,\,\, k=i \\
-\gamma_k & {\rm otherwise} \quad .
\end{array}
\right.
\end{equation}

\end{itemize}

\subsection{Winding trajectories}

We start with the case that all vortices move along closed
trajectories. If $m_k$ is the
number of windings of the $k$th vortex around other vortices, then
$\Omega_k$ changes by
$2\pi m_k$. In view of Eqs.\ (\ref{zeromodespinor}) and (\ref{gamma}),
this implies that
the Majorana operator $\gamma_k$ is multiplied by $(-1)^{m_k}$.

In the simplest case vortex $1$ encircles vortex $2$, leading to both
$\gamma_1$ and
$\gamma_2$ being multiplied by $-1$, with all other operators
unchanged. This is a
consequence of the fact that a phase factor $2\pi$ of the order
parameter leads to a
phase $\pi$ (and thus a minus sign) for the fermionic operators.

In the general case, we need to find a unitary transformation
$U_{\rm wind}$ by
requiring
\begin{equation}
U_{\rm wind} \gamma_k U_{\rm wind}^\dagger=\left\{
\begin{array}{cc}
-\gamma_k & { \rm when }\,\,\,\,\,m_{k}\,\,\,\,\,\,\hbox{\rm is odd} \\
\gamma_k  & { \rm when }\,\,\,\,\,m_{k}\,\,\,\,\,\hbox{\rm is even}
\end{array}
\right.\quad .  \label{udef}
\end{equation}
This is satisfied by the operator
\begin{equation}
   U_{\rm wind}=\prod_{k=1}^{2N}u_{k}^{m_{k}},  \label{uwinding}
\end{equation}
where the operator ordering in $U_{\rm wind}$ is not important up to an
overall minus
sign.

\subsection{Exchange trajectories}

Exchange trajectories in which some of the vortices trade places are
more complicated
since the phase changes of $\Omega_{k}$ associated with a particular
trajectory do not
only depend on the winding numbers, but also on the details of the
trajectory and on the
precise definition of the cut of the function $\arg ({\bf r})$ where its
value jumps by
$2\pi$.

The simplest example is the interchange of two vortices. Inevitably one
of the vortices
crosses the cut line of the other vortex. Thus such an interchange
involves an operator
$u_{12}$ that exchanges the positions of the two vortices and an
operator $ u_{1}$ or
$u_{2}$ that multiplies the appropriate vortex by $-1$. As a result, the unitary
transformations
generated by these exchanges are
\begin{eqnarray}
U^{1\leftrightharpoons 2} &=&\frac{1}{\sqrt{2}}\left( 1+u_{2}u_{1}
\right) \\
U^{1\rightleftharpoons 2} &=&\frac{1}{\sqrt{2}}\left( 1+u_{1}u_{2}
\right).
\end{eqnarray}
The transformation $U^{1\leftrightharpoons 2}$ transforms $c_{1}
\rightarrow c_{2}$ and
$c_{2}\rightarrow -c_{1}$ while $U^{1\rightleftharpoons 2}$ transforms $c_{1}
\rightarrow
-c_{2}$ and $ c_{2}\rightarrow c_{1}$.

The combination of the motion of the first vortex from ${\bf R}_{1}$ to ${\bf R}_{2}
$ and the motion
of the second vortex from ${\bf R}_{2}$ to ${\bf R}_{1}$ generates a closed curve.
When that curve
encloses a third vortex, the phase of that vortex changes by $2\pi $,
and so does the
phase of one of the first two vortices. Altogether, this gives three
phase shifts of
$2\pi$, and the transformations $U^{1\leftrightharpoons 2}$ and $U^{1
\rightleftharpoons
2}$ have to be multiplied by either $u_{3}u_{1}$ or $u_{3}u_{2}$. There
are two possible
outcomes to that. The first is $u_{3}\left( u_{1}+u_{2}\right)$, a
transformation in
which the phase of each of the three vortices is shifted by $2\pi$. The
second is
$u_{3}\left( u_{1}-u_{2}\right)$, in which the phase of the third
vortex is shifted by
$2\pi$, and the other two $2\pi$ shifts are both given to one vortex.

\section{Orthogonality of vortex states}
\label{orthogonal}

In this appendix, we discuss the scheme to generate the functions $w^
{(k)}_i( {\bf r})$
in more detail and prove the orthogonality relation Eq.\ (\ref
{wortho}). We start by
proving the orthogonality of the states $w_i^{(0)}({\bf r})$ and $w_j^
{(1)}({\bf r})$. In
spinor notation, we can write
\begin{equation}
    Z_i^{(1)} = i\sum_E {\rm sgn} E \,\, C^i_E \left(u_E({\bf r})\atop
v_E({\bf r})\right)
         ={1\over \sqrt{2}}\left(w_i^{(1)}({\bf r})\atop [w_i^{(1)}
({\bf r})]^*\right)
\end{equation}
in terms of the expansion coefficients
\begin{equation}
C_E^{j} ={1\over\sqrt{2}} \left\langle {u_E \atop v_E} \left| { iw_{j}^
{(0)} \atop
-i[w_{j}^{(0)}]^*}\right.\right\rangle .
\end{equation}
These coefficients satisfy the relation
\begin{equation}
C_{-E}^{j} = [ C_E^{j}]^*  \label{coef1}
\end{equation}
whose proof uses the identity $u_E({\bf r}) =[v_{-E}({\bf r} )]^*$.
The states $w_i^{(0)}({\bf
r})$ and
$w_j^{(1)}({\bf r})$ are obviously orthogonal due to their
localization
properties for $i\neq j$. For $i=j$, the orthogonality follows from
\begin{equation}
{\rm Re} \langle w_i^{(1)}| w_i^{(0)}\rangle = {1\over 2}\left\langle {
w_i^{(1)} \atop
[w_i^{(1)}]^*}\left| { w_i^{(0)} \atop [w_i^{(0)}]^* }
\right.\right\rangle =0
\label{reortho1}
\end{equation}
and
\begin{equation}
{\rm Im} \langle w_i^{(1)}| w_i^{(0)}\rangle = -{1\over 2} \left\langle
{ w_i^{(1)} \atop
[w_i^{(1)}]^*} \left| { iw_i^{(0)} \atop -i[w_i^{(0)}]^* }
\right.\right\rangle ={i\over
2}\sum_E {\rm sgn}E\,\,|C_E^{i}|^2 =0. \label{imortho1}
\end{equation}
Here, Eq.\ (\ref{reortho1}) uses that by construction, the spinor $Z_i^
{(1)}$ has zero
overlap with the zero-energy spinors. Eq.\ (\ref{imortho1}) uses the
symmetry
(\ref{coef1}) of the expansion coefficients $C_E^{i}$.

We now turn to the higher iterations $w_i^{(k)}({\bf r})$ and proceed to prove the orthogonality Eq.\ (\ref{wortho}).
Using the completeness of the eigenfunctions of the BdG equations, we
obtain the useful
result
\begin{equation}
S^\dagger S = {\bf 1} -{1\over 2} \sum_j \left|{w_j^{(0)} \atop
[w_j^{(0)}]^*}\right\rangle \left\langle{w_j^{(0)} \atop [w_j^{(0)}]^*}
\right|\; ,
\end{equation}
with $S$ a hermitian operator, $S=S^\dagger$.

For different vortices,
$i\neq j$, the
orthogonality follows again from the locality properties of the $w$'s. Thus,
we focus on states from the same vortex, $i=j$,
and drop the subscript labelling the vortex in the remainder of this
appendix. In each
iterative step $k$ of the construction of the $w$'s, we need to prove
that
\begin{equation}
\langle w^{(l)}|w^{(k)}\rangle=0
\end{equation}
for all $l<k$. Since this amounts to a proof by induction, we may
exploit that two states
$w$ with indices smaller than $k$ are orthogonal.

Starting with $l=0$, we have
\begin{equation}
2{\rm Re}\langle w^{(k)}|w^{(0)}\rangle = \left\langle {w^{(k)} \atop
[w^{(k)}]^*}\left|
{w^{(0)} \atop [w^{(0)}]^*}\right.\right\rangle =0
\end{equation}
since by construction, the spinor $Z_k$ has zero overlap with zero-energy spinors.
Furthermore,
\begin{eqnarray}
-2{\rm Im}\langle w^{(k)}|w^{(0)}\rangle &=& \left\langle {w^{(k)}
\atop
[w^{(k)}]^*}\left| {iw^{(0)} \atop -i[w^{(0)}]^*}\right.\right\rangle =-
i\left\langle
{w^{k-1} \atop [w^{k-1}]^*}\right| \sigma_z S \sigma_z \left|{w^{(0)}
\atop [w^{(0)}]^*}
\right\rangle  \nonumber \\
&=&i\left\langle {w^{k-1} \atop [w^{k-1}]^*}\right| \sigma_z \left|{ w^
{(1)} \atop
[w^{(1)}]^*}\right\rangle=0\; .
\end{eqnarray}
For $l=1,2\ldots k-2$, we obtain
\begin{eqnarray}
2{\rm Re}\langle w^{(k)}|w^{(l)}\rangle &=& \left\langle {w^{(k)} \atop
[w^{(k)}]^*}\left| {w^{(l)} \atop [w^{(l)}]^*}\right.\right\rangle
=\left\langle
{w^{(k-1)} \atop [w^{(k-1)}]^*}\right| (-S\sigma_z)^\dagger(-S\sigma_z)
\left|{w^{(l-1)}
\atop [w^{(l-1)}]^*}
\right\rangle  \nonumber \\
&=& \left\langle {w^{(k-1)} \atop [w^{(k-1)}]^*}\left| {w^{(l-1)} \atop [w^{(l-1)}]^*}
\right.\right\rangle \nonumber \\
&&\,\,\,\,\,\,\,\,
-{1\over 2}\left\langle {w^{(k-1)} \atop [w^{(k- 1)}]^*} \right|
\sigma_z \left|{w^{(0)} \atop [w^{(0)}]^*}\right\rangle \left\langle {w^ {(0)} \atop
[w^{(0)}]^*}\right|\sigma_z \left|{w^{(l-1)} \atop [w^{(l-1)}]^*} \right\rangle =0
\label{reorthoi}
\end{eqnarray}
and
\begin{eqnarray}
-2{\rm Im}\langle w^{(k)}|w^{(l)}\rangle &=& \left\langle {w^{(k)}
\atop
[w^{(k)}]^*}\left| {iw^{(l)} \atop -i[w^{(l)}]^*}\right.\right\rangle
= -i\left\langle
{w^{(k-1)}\atop [w^{(k-1)}]^*}\right| \sigma_z S \sigma_z \left|{w^
{(l)} \atop
[w^{(l)}]^*}
\right\rangle  \nonumber \\
&=& i\left\langle {w^{(k-1)} \atop [w^{(k-1)}]^*}\right| \sigma_z \left|
{ w^{(l+1)}\atop
[w^{(l+1)}]^*}\right\rangle = \left\langle {w^{(k-1)} \atop [w^{(k-1)}]
^*} \left|
{iw^{(l+1)} \atop -i[w^{(l+1)}]^*}\right.\right\rangle=0\; .
\end{eqnarray}
Finally, we note for $l=k-1$ that ${\rm Re}\langle w^{(k)}|w^{(k-1)}
\rangle=0$ can be
obtained in complete analogy to Eq.\ (\ref{reorthoi}) and
${\rm Im}\langle w^{(k)}|w^{(k-1)}\rangle =0$ in analogy with Eq.\ (\ref{imortho1}).
This completes the iterative construction of the functions $w^{(k)}_j$
including the proof of
the orthogonality relations Eq.\ (\ref{wortho}).


\begin{thebibliography}{50}
\bibitem{foot1} Here, the convention of where to place the $i$ is
  chosen different from Eq.\ (\ref{alpha}) to minimize the number of
  explicit factors of $i$ appearing in subsequent equations.
\bibitem{foot2} The product is taken to run from left to right in
  descending order of the index $2j$.
\bibitem{foot4} The same is true for all generations; however, for all
  generations but the ``last" one, this is guaranteed by
  $Y_i^{(k)}|{\rm gs}\rangle = 0$ since $X_i^{(k-1)} =
  (1/\sqrt{2})\{Y_i^{(k)}+[Y^{(k)}_i]^\dagger\}$.
\end{thebibliography}
\end{document}